\documentclass[aps,prl,twocolumn,superscriptaddress,longbibliography]{revtex4-2}

\usepackage{amsmath,amssymb,bm}
\usepackage{graphicx}
\usepackage{xcolor}
\usepackage{ulem}

\begin{document}

\title{From GeV to PeV: Orbital Modulation and Wind-Regulated Ultrahigh-energy Emission in Cygnus~X-3}

\author{Su-Yu Wan}
\affiliation{School of Astronomy and Space Science, Nanjing University, Nanjing 210023, China}
\affiliation{Key Laboratory of Modern Astronomy and Astrophysics (Nanjing University), Ministry of Education, Nanjing 210023, China}

\author{Ruo-Yu Liu}
\email{ryliu@nju.edu.cn}
\affiliation{School of Astronomy and Space Science, Nanjing University, Nanjing 210023, China}
\affiliation{Key Laboratory of Modern Astronomy and Astrophysics (Nanjing University), Ministry of Education, Nanjing 210023, China}
\affiliation{Tianfu Cosmic Ray Research Center, Chengdu 610000, Sichuan, China}

\author{Xing-Fu Zhang}
\affiliation{School of Astronomy and Space Science, Nanjing University, Nanjing 210023, China}
\affiliation{Key Laboratory of Modern Astronomy and Astrophysics (Nanjing University), Ministry of Education, Nanjing 210023, China}

\author{Dmitriy Khangulyan}
\affiliation{Tianfu Cosmic Ray Research Center, Chengdu 610000, Sichuan, China}
\affiliation{State Key Laboratory of Particle Astrophysics, Institute of High Energy Physics, Chinese Academy of Sciences, Beijing 100049, China}

\begin{abstract}
Recent LHAASO observations reveal ultrahigh-energy (UHE) gamma-ray emission from Cygnus~X-3 with evidence for orbital modulation, providing a rare opportunity to study particle acceleration and transport on binary scales. We develop a common binary--jet model that jointly explains the contemporaneous GeV--PeV spectra and orbital light curves. The GeV emission is produced by anisotropic inverse-Compton scattering of stellar photons by relativistic electrons, while multi-PeV protons escaping from the inner region produce TeV--PeV gamma rays through photohadronic and hadronuclear interactions. The GeV orbital modulation constrains the binary--jet geometry, allowing the UHE observations to probe the escaping proton population and its interaction with the circum-binary environment. We find that the UHE emission is strongly regulated by the accretion-driven wind structure. In the present state, a fast disk wind displaces much of the denser Wolf--Rayet (WR) wind from the escaping-proton trajectories, suppressing hadronuclear emission. Changes in the strength, collimation, or orientation of the disk wind can expose the escaping protons to a much denser WR wind and enhance the PeV gamma-ray luminosity by orders of magnitude. This provides a possible connection to the much brighter, though unconfirmed, PeV activity reported in the 1980s, and links UHE gamma-ray activity to the accretion and outflow state of the binary.
\end{abstract}

\maketitle

Cygnus~X-3 is a compact X-ray binary with a 4.8\,hr orbital period \cite{vanKerkwijk1992,Zdziarski2013,Giacconi1967}, composed of a compact object with relativistic jets and a Wolf--Rayet (WR) star. Transient GeV emission detected by AGILE and \textit{Fermi}-LAT is orbitally modulated \cite{Tavani2009,Abdo2009}. More recently, LHAASO detected variable ultrahigh-energy gamma-ray emission ($>100\,$TeV) from Cygnus~X-3 up to $3.7$~PeV, contemporaneous with a GeV high state and showing evidence for modulation at the same 4.8~hr period \cite{Cao2026_CygX3}. The orbital variability places the UHE emission site within or close to the binary system. Given such a compact origin, a leptonic interpretation is strongly disfavored, implying proton acceleration to tens of PeV. It makes Cygnus~X-3 the first confirmed Galactic super-PeVatron and a special laboratory for studying multi-PeV particle acceleration and transport. Interestingly, several air-shower experiments in the 1980s also reported orbitally modulated emission at PeV energies, with fluxes orders of magnitude above those measured today \cite{Samorski1983,LloydEvans1983,Kifune1986,Alexeenko1987}, although these claims were not confirmed by later high-statistics
searches \cite{Borione1997}.

Several recent studies have addressed different aspects of LHAASO observations. The LHAASO spectrum can be reproduced by photohadronic interactions in the intense stellar radiation field \cite{Cao2026_CygX3}, while both photohadronic and hadronuclear channels may contribute under different conditions in compact outflows \cite{Wei2025_MQcore,2026NewAR.10201749H}. Ref.\cite{Zdziarski2026_PeV} considered high-energy protons escaping from a tilted jet and interacting with the stellar radiation field and WR wind, reproducing the TeV--PeV spectrum and orbital modulation. At lower energies, \citet{Zhang2026_GeV} modeled the contemporaneous \textit{Fermi}-LAT spectrum and GeV light curve with anisotropic external inverse-Compton emission. They constrained the GeV emitter to a distance of a few $10^{11}$~cm from the compact object and showed that this region cannot confine the multi-PeV protons required by the LHAASO emission, suggesting that multi-PeV protons must escape the jets before being advected to the GeV emission zone. The GeV and UHE light curves also peak at different orbital phases \cite{Zhang2026_GeV}, further indicating that the two bands probe different emission sites within the same binary system.

Despite these efforts, the spectral and temporal information across the GeV--PeV band has not yet been explained simultaneously within a single physical framework. This limits what can be learned from the UHE emission alone, because both its spectrum and orbital modulation depend on the injected proton population, the system geometry, and the radiation and gas fields encountered by the escaping particles. The high-statistics GeV data provide an independent and powerful constraint on the geometry. As the GeV emission is produced by anisotropic inverse-Compton scattering of the stellar radiation, its orbital modulation is highly sensitive to the orientation of the jet relative to the star and the observer. Once the binary--jet geometry is constrained in this way, the UHE spectrum and orbital modulation can be used more directly to probe where multi-PeV protons are released into the binary environment and how they subsequently propagate through it.

In this Letter, we develop a unified model for the GeV--PeV emission of Cygnus~X-3 that exploits these complementary constraints. The GeV component is produced by anisotropic inverse-Compton scattering of stellar photons by relativistic electrons in the jet, while multi-PeV protons released from the inner region propagate through the binary and produce TeV--PeV gamma rays via $p\gamma$ and $pp$ interactions. We account for the interaction between the accretion-driven disk wind and the dense WR wind, which controls the target gas column for the escaping particles. We show that this framework can reproduce the contemporaneous GeV--PeV observations and that changes in the circum-binary wind configuration can alter the hadronic gamma-ray output by orders of magnitude, providing a possible connection to the much brighter PeV emission reported in the 1980s.

\begin{figure}[htbp]
\centering
\includegraphics[width=0.98\columnwidth,trim=10.0cm 0cm 0cm 0cm,clip]{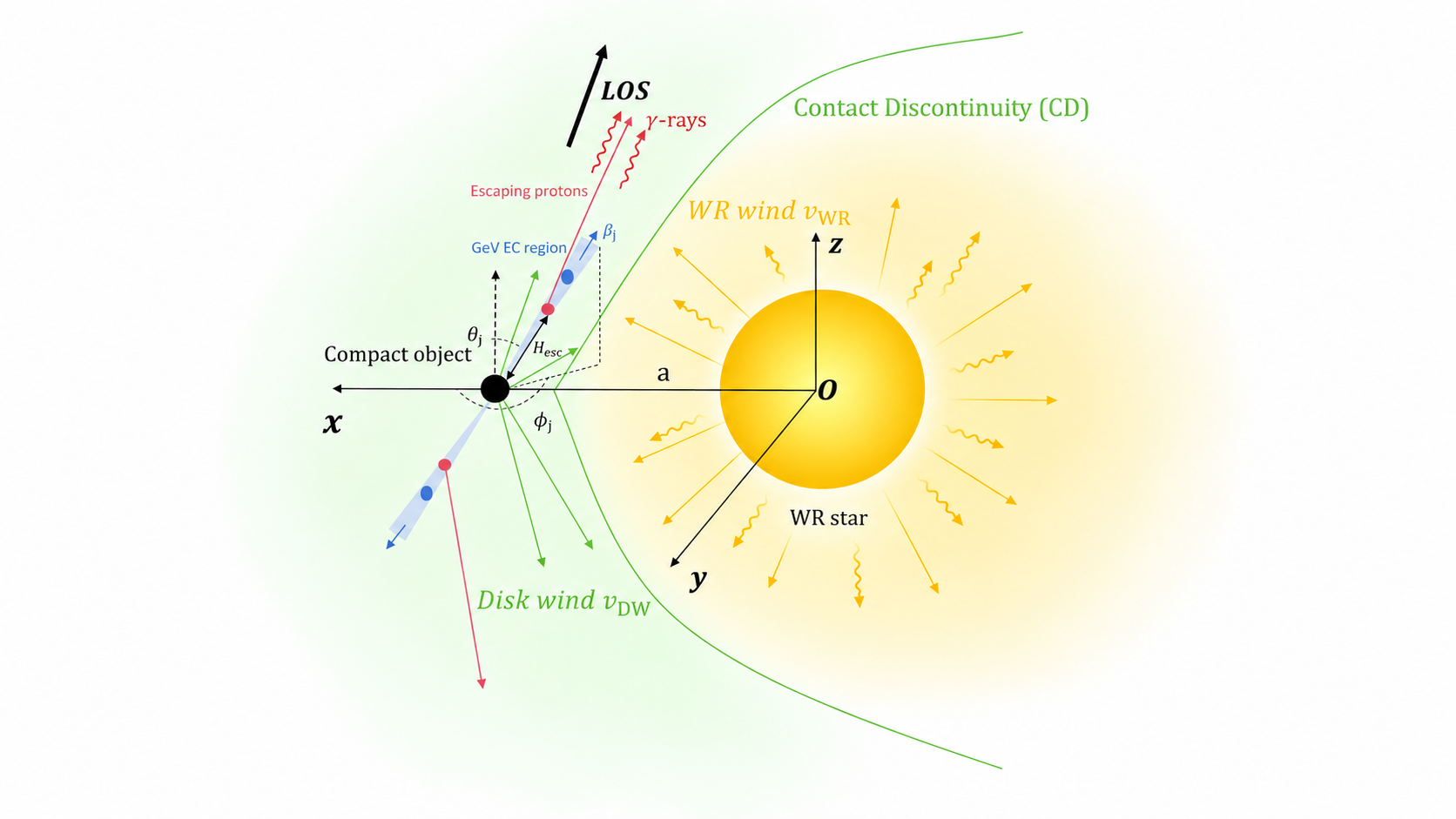}
\caption{Sketch of binary--jet geometry. GeV emission is produced at $H_{\rm GeV}$, while high-energy protons are released at an effective height $H_{\rm esc}$ and interact with stellar photons and circum-binary gas. The WR and disk winds are separated by a contact discontinuity.}
\label{fig:sketch}
\end{figure}

\textit{Unified model.--}
We use a common binary--jet geometry for both components, illustrated in Fig.~\ref{fig:sketch}. The WR star is at the origin, the orbit lies in the $x$--$y$ plane, and phase $\phi=0$ corresponds to superior conjunction of the compact object. The separation is $a=2.65\times10^{11}\,$cm and the observer inclination is fixed to $i=30^\circ$ \cite{Antokhin2022}. The jet direction is specified by $(\theta_{\rm j},\varphi_{\rm j})$ and is shared by the GeV and PeV calculations.

The GeV component is produced by anisotropic external IC
scattering (EC) of stellar photons by relativistic electrons confined in the jet at a characteristic height $H_{\rm GeV}$, whose injection is parameterized as $Q_e'(\gamma
')\propto K_{\rm inj}\gamma'{}^{-p_e}$. Here the primed quantities are defined in the jet comoving frame. Because the scattering geometry changes with orbital phase, the GeV light curve is particularly sensitive to the jet orientation, velocity $\beta_{\rm j}$, and emitter location. In the joint fit, the much higher GeV statistics provide the primary
constraint on the binary--jet geometry. For the UHE component, we do not specify the microscopic acceleration mechanism. Instead, $H_{\rm esc}$ parameterizes the location at which high-energy protons are released from the jets into the circum-binary environment. The intrinsic escaping proton spectrum is assumed to be isotropic in the jet comoving frame and is parameterized as
\begin{equation}
Q'_p(E'_p, \Omega^{\prime})=K'_p E_p'^{-s_p}\exp \left(-\frac{E'_p}{E'_{p,\rm cut}}\right)\,,
\label{eq:qp}
\end{equation}
with equal intrinsic proton power $L_{\rm p}^{\prime}/2$ assigned to the jet and counterjet. For the jet/counterjet branch $\xi=\pm1$, the corresponding differential spectrum of injected protons is anisotropic in the observer frame: $Q_{p,\xi}(E_p,\Omega)= \Gamma_{\rm j}\delta_\xi Q'_p\left(E_p/\delta_\xi, \Omega^{\prime}\right)$, where $\delta_\xi(\Omega)=\left[\Gamma_{\rm j}\left(1-\xi\beta_{\rm j}\boldsymbol e_{\rm j}\cdot\boldsymbol e_p\right)\right]^{-1}$ represents the Doppler factor toward the proton propagation direction $\boldsymbol e_p$ (see Appendix~\ref{app:spec} for the derivation). For multi-PeV protons, the Larmor radius in the circum-binary magnetic field can be comparable to or larger than the binary scale provided the magnetic field does not exceed \(\sim10\,{\rm G}\), motivating a ballistic treatment after escape. Because the secondary gamma rays from multi-PeV proton interactions are strongly beamed along the parent-proton direction, the observed emission is dominated by protons propagating toward the observer.

Along each trajectory, the protons interact with the stellar radiation field and with circum-binary gas. For the jet/counterjet branch, the compact object is at $\boldsymbol r_c(\phi)$ at orbital phase $\phi$, and the observer-directed protons follow
\begin{equation}\label{eq:traj}
\boldsymbol r_{p,\xi}(l,\phi)=\boldsymbol r_c(\phi)+\xi H_{\rm esc}\boldsymbol e_j+l\boldsymbol e_{\rm obs}.
\end{equation}
The $p\gamma$ component is then calculated as
\begin{equation}\label{eq:pgamma}
\begin{split}
\frac{dN_{\gamma,\xi}^{p\gamma}}{dE_\gamma dt}(E_\gamma,\phi)=
&\int dE_p\,Q_{p,\xi}(E_p,\Omega_{\rm obs} )\int dl\int d\epsilon\,d\Omega_\star\, n_\star(\epsilon,\Omega_\star,\boldsymbol r_{p,\xi})\\
&\times (1-\mu_{p,\xi})\, \sigma_{p\gamma}(\eta_\xi)\frac{dN_\gamma^{p\gamma}}{dE_\gamma}(\eta_\xi,x),
\end{split}
\end{equation}
where $\mu_{p,\xi}$ is the cosine of the proton--photon collision angle, and $\delta_\xi(\Omega_{\rm obs})=\left[\Gamma_{\rm j}\left(1-\xi\beta_{\rm j}\boldsymbol e_{\rm j}\cdot\boldsymbol e_{\rm obs}\right)\right]^{-1}$ . The two dimensionless parameters are $\eta_{\xi}=2E_p\epsilon(1-\mu_{p,\xi})/(m_p^2c^4)$ and $x=E_\gamma/E_p$. Both the stellar photon density and the proton--photon collision angle vary with orbital phase. 

The $pp$ component from the same escaping proton population is
\begin{equation}\label{eq:pp}
\begin{split}
\frac{dN_{\gamma,\xi}^{pp}}{dE_\gamma dt}(E_\gamma,\phi)=
&\int dE_p\,Q_{p,\xi}(E_p, \Omega_{\rm obs})\int dl\,n_{\rm gas}(\boldsymbol r_{p,\xi})\sigma_{pp}(E_p) \\
&\times \frac{dN_\gamma^{pp}}{dE_\gamma}(E_\gamma,E_p),
\end{split}
\end{equation}
where the phase dependence of $n_{\rm gas}$ reflects both the proton trajectory and the gas distribution. The $p\gamma$ yield $dN_\gamma^{p\gamma}/{dE_\gamma}$ is calculated with \texttt{SOPHIA++}, while the $pp$ yield $dN_\gamma^{pp}/{dE_\gamma}$ follows Ref.~\cite{Kelner2006}. 
Eq.~(\ref{eq:pgamma}) shows explicitly why the $p\gamma$ modulation is sensitive to the binary geometry, whereas Eq.~(\ref{eq:pp}) shows that the hadronuclear emission is controlled mainly by the gas column. Details of the  calculation such as the gas and radiation density distribution are given in the Appendix.

The gas distribution is set by the interaction between the WR wind and an accretion-driven disk wind. We adopt $\dot M_{\rm WR}=5\times10^{-6}M_\odot\,{\rm yr^{-1}}$ and $v_{\rm WR}=1000\,{\rm km\,s^{-1}}$ for the WR wind \cite{Koljonen2017}, and use $\dot M_{\rm DW}=10^{-6}M_\odot\,{\rm yr^{-1}}$ as a characteristic disk-wind mass-loss rate. The disk-wind velocity $\beta_{\rm DW}$ determines the size of the funnel around the jet and is treated as a free parameter. The two winds are separated by a contact discontinuity obtained from ram-pressure balance \cite{Canto1996}. Because the disk wind is much faster, its density can be about two orders of magnitude below that of the WR wind at comparable distances even when its momentum flux dominates the wind interaction.

\begin{figure*}[htbp]
\centering
\includegraphics[width=0.94\textwidth]{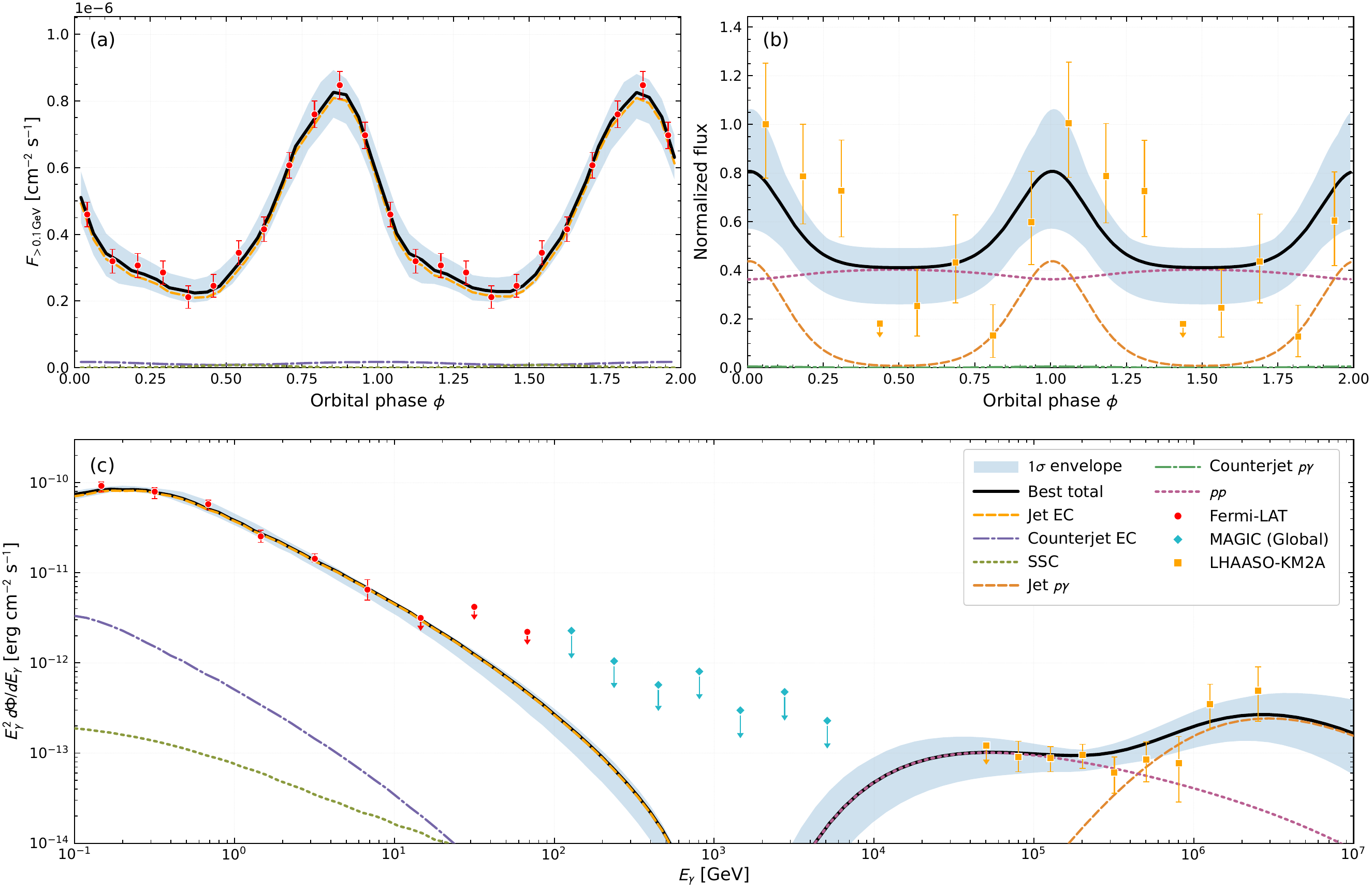}
\caption{Joint fit to the GeV--PeV spectra and orbital modulation of Cygnus X-3. The black solid curves show the total best-fit models, and the blue shaded regions denote the corresponding $1\sigma$ statistical uncertainties. The orange dashed, purple dash-dotted, and green dotted lines represent the jet External-Compton (EC), counterjet EC and Synchrotron Self-Compton (SSC), respectively, while the orange dashed, green dash-dotted, and magenta dotted curves at UHE energies denote the $p\gamma$ component from protons in the approaching jet, the $p\gamma$ component from protons in the counterjet, and total $pp$ components. (a) GeV gamma-ray orbital light curves. Red circles show the Fermi-LAT data during the LHAASO detection period \citep{Cao2026_CygX3}. (b) UHE gamma-ray orbital light curve, with LHAASO data shown in orange squares. The normalization of the UHE light curve is fixed at $1.053\times 10^{-15}\rm\, cm^{-2}s^{-1}$ following \citet{Cao2026_CygX3}. The $\gamma\gamma$ absorption in the stellar radiation field, along with the attenuation caused by interstellar radiation field (ISRF) and Cosmic Microwave Background (CMB) photons are all included. (c) GeV--PeV Spectral Energy Distribution (SED). Red circles and orange squares represent the \textit{Fermi}-LAT and LHAASO data respectively, while cyan diamonds denote the upper limits given by MAGIC \citep{2026JHEAp..5300630A}. Both the data and the model have been corrected for the attenuation by CMB and ISRF.}
\label{fig:joint}
\end{figure*}

\textit{Joint spectral and orbital constraints.--}
The joint fit reproduces all four observables (i.e., spectra and orbital light curves of \textit{Fermi}-LAT and LHAASO), as shown in Fig.~\ref{fig:joint}. As expected, the geometry is constrained primarily by the GeV data,
which give \(H_{\rm GeV}=2.75^{+0.33}_{-0.60}\times10^{11}\,{\rm cm}\), \(\beta_{\rm j}=0.603^{+0.076}_{-0.071}\), \(\theta_{\rm j}=45.0^{+10.7}_{-3.5}\) deg, and \(\varphi_{\rm j}=196.3^{+6.1}_{-2.3}\) deg, consistent with Ref.~\cite{Zhang2026_GeV}. Under such a geometry, the model favors proton release on the sub-binary scale, with a height $H_{\rm esc}\sim3.0\times10^{10}\,{\rm cm}$, noting that the present LHAASO statistics do not exclude release close to the compact object.
The UHE spectrum constrains the escaping proton population to $s_p=2.461^{+0.213}_{-0.165}$ and $E'_{p,\rm cut}=108^{+440}_{-80}\,{\rm PeV}$. The intrinsic proton luminosity is $L'_p(>1\,{\rm TeV})=0.583^{+2.574}_{-0.355}\times10^{38}\,{\rm erg\,s^{-1}}$, while the differential power at 1\,PeV in the observer frame is $E_p^2\sum_{\xi=\pm1}\int d\Omega\,Q_{p,\xi}(1 \, \rm PeV,\Omega)=2.02^{+2.78}_{-0.75}\times10^{36}\,{\rm erg\,s^{-1}}$. In the preferred wind geometry, a relatively flat $pp$ contribution is important below a few 100\,TeV, whereas a bump-like $p\gamma$ component dominates near 1\,PeV and controls most of the orbital modulation. A phenomenological low-energy suppression $\exp(-10\,{\rm TeV}/E)$ is applied to the obtained gamma-ray spectrum, representing the inefficient escape of lower-energy protons, which may lose energy adiabatically. It has little influence on the inferred parameters.

\textit{Wind-regulated PeV activity.--}
The UHE gamma-ray luminosity is regulated by both the power of escaping protons and the efficiency with which that power is converted into gamma rays. In our model, the latter is strongly controlled by the circum-binary gas distribution. The fast disk wind displaces the much denser WR wind from a large fraction of the trajectories sampled by the observer-directed protons, thereby suppressing the $pp$ emission. Such an environmental dependence of the $pp$ component provides a possible connection to the historical PeV claims. If the wide-angle disk wind were suppressed, became more collimated, or changed orientation, escaping protons could propagate mainly through the denser WR wind, substantially enhancing the hadronuclear interaction efficiency. Indeed, disk winds in black-hole X-ray binaries are state dependent \cite{Miller2012,ProgaKallman2002}. In Cygnus~X-3, radio imaging also shows substantial changes in jet position angle between outbursts, consistent with jet precession \cite{Mioduszewski2001,MillerJones2009}. If the disk wind follows the inner-flow or jet axis, such reorientation can move the disk-wind cone away from the observer-directed proton trajectories without requiring the accretion power or proton accelerator to switch off.

\begin{figure*}[t]
\centering
\includegraphics[width=0.94\textwidth]{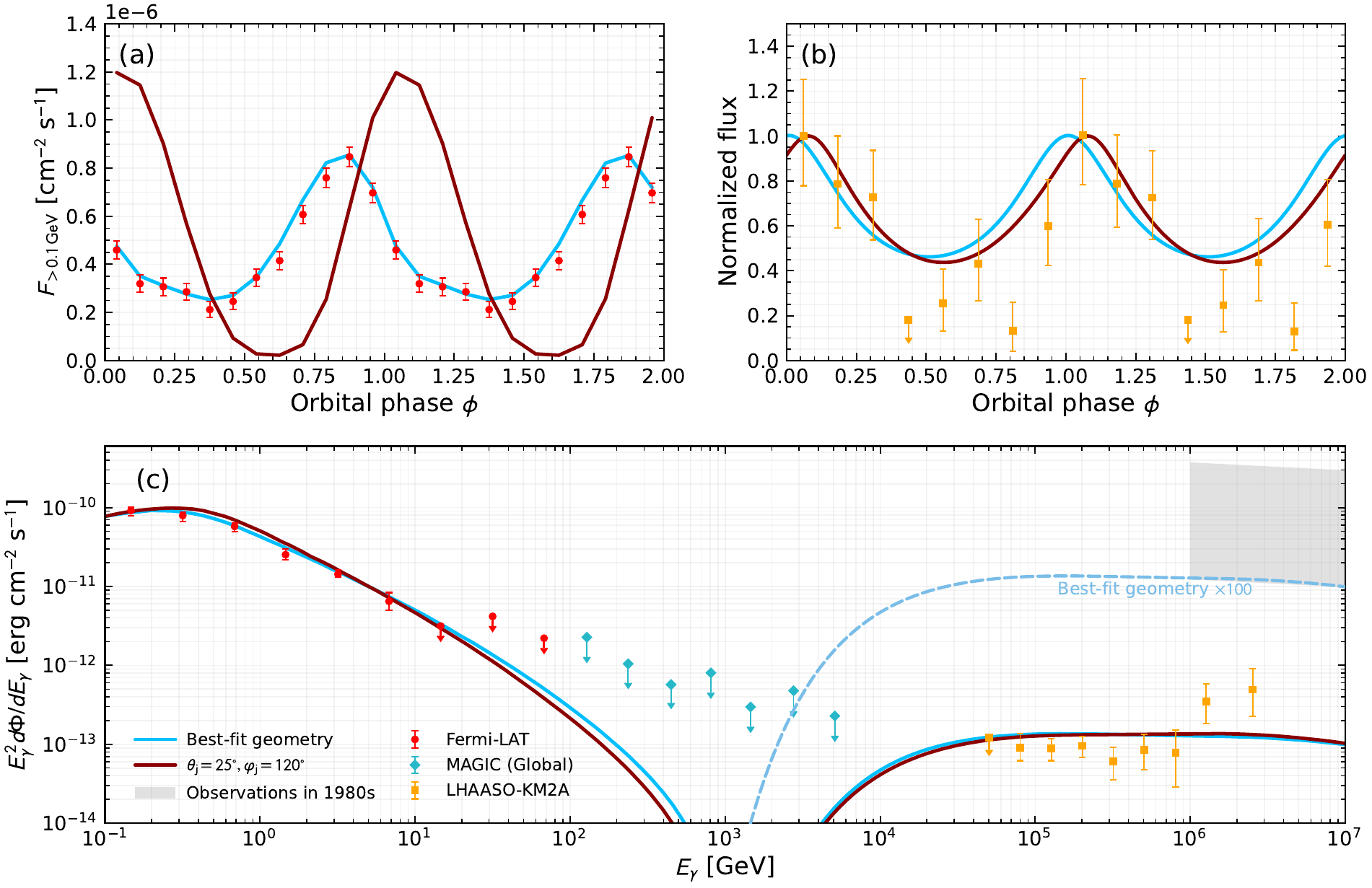}
\caption{Same as Fig.~\(\ref{fig:joint}\) but for the pure WR-wind scenario with two configurations of the jet geometry and escaping particle distribution. The blue solid curves adopt our best-fit jet geometry, including the relativistic transformation of the escaping-proton distribution. The maroon curves adopt \(\theta_{\rm j}=25^\circ\), \(\varphi_{\rm j}=120^\circ\), and \(H_{\rm esc}=0.7a\) following \citet{Zdziarski2026_PeV}, with an isotropic distribution of escaping protons in the observer frame, serving as a cross check of our calculation. Both cases use \(L_{p}(>1\,{\rm TeV})\simeq2\times10^{36}\,{\rm erg\,s^{-1}}\) and a proton spectral index \(s_p=2.2\), without a high-energy exponential cutoff to facilitate a direct comparison. A smooth low-energy exponential cutoff is also adopted, as in Fig.~\(\ref{fig:joint}\). In panel (c), the rough flux range reported by air-shower experiments in the 1980s \citep{Samorski1983,LloydEvans1983,Kifune1986,Alexeenko1987} is shown as a gray band. The blue dashed curve shows the flux under the best-fit geometry by raising the proton power to the same level as in Fig.~\(\ref{fig:joint}\), i.e., \(L_p(>1\,{\rm TeV})\simeq2\times10^{38}\,{\rm erg\,s^{-1}}\).}
\label{fig:historical}
\end{figure*}

A pure WR-wind scenario can increase the target column and hence the $pp$ luminosity by roughly two orders of magnitude. For a comparable proton power, the resulting PeV flux reaches the level reported by the early air-shower experiments (Fig.~\ref{fig:historical}). Nevertheless, the historical detections remain uncertain, so future long-term PeV monitoring, together with measurements of the accretion and wind states of Cygnus~X-3, will be crucial to test this possibility. 

\textit{Implications.--}
The fitted escaping proton power provides a direct scale for the contribution of Cygnus X-3 to Galactic PeV cosmic rays. A standard steady-state diffusion estimate requires $E_p^2Q_{p,\rm Gal}(1\,\mathrm{PeV})\sim4\times10^{38}(D_{\rm PeV}/4\times10^{30}\,\mathrm{cm^2\,s^{-1}})\,\mathrm{erg\,s^{-1}}$ per logarithmic interval to maintain the Galactic PeV CR flux \cite{Strong2007,Lipari2014}. Thus even the gamma-ray high state supplies only a small fraction of the Galactic requirement, and the long-term fraction is smaller if PeV acceleration is intermittent. Current gamma-ray data suggest an active-state fraction of order $0.1$--$0.4$\cite{Prokhorov2023,Cao2026_CygX3}, although this should not be regarded strictly as the proton injection duty cycle over the residence timescale of CRs.
The result also constrains the recent proposal that Cygnus~X-3 powers the highest-energy emission of the Cygnus Bubble \cite{Shi2026_CygBubble}. The model requires an injection power corresponding to $E_p^2Q_p(1\,\mathrm{PeV})\simeq(4.7$--$9.3)\times10^{36}\,\mathrm{erg\,s^{-1}}$ for an $E^{-2}$ spectrum with a 10 PeV cutoff, maintained for $\sim 100-400\,$kyr. This is marginally consistent with our result for high states, so a dominant contribution to the Bubble would require a larger time-averaged proton power or high duty cycle.

In summary, we have shown that the contemporaneous GeV and UHE gamma-ray spectra and orbital modulation of Cygnus~X-3 can be described within a common binary--jet geometry. The UHE emission is produced by an escaping population of multi-PeV protons. The UHE gamma-ray luminosity is regulated by the circum-binary wind structure. Changes in the disk-wind configuration can modify the luminosity by orders of magnitude and provide a possible connection between the present LHAASO state and the much brighter, though unconfirmed, PeV activity of this object reported in the 1980s. Our results suggest a connection between UHE gamma-ray activity with the accretion/outflow state of the binary. Joint long-term multiwavelength observations may test this picture. If the significant change in the UHE luminosity is associated with a reorientation of the jet and wide-angle disk wind, corresponding changes in the Fermi-LAT orbital light curve would be expected. Contemporaneous X-ray observations may trace the accretion and wind state, while radio imaging can probe changes in the jet orientation and ejecta. It may then establish UHE gamma rays as a probe of accretion-driven outflows in compact binaries, and offer a new insight into the interplay between black-hole accretion and extreme particle acceleration.

\begin{acknowledgments}
This work is supported by the National Natural Science Foundation of China under Grants No.~12593852 and No.~12333006, and by the Basic Research Program of Jiangsu under Grant No.~BK20250059. DK acknowledges support by RSF grant No.~24-12-0045
\end{acknowledgments}

\nocite{Mucke2000,ABR2022,1967PhRv..155.1404G,1967PhRv..155.1408G,2009ApJ...707..916F,2017MNRAS.470.2539P,2007ARA&A..45..177C,Koljonen2017,Vilhu2021,King2016}
\bibliography{ms}

\newpage
\appendix

\section{Spectrum of escaping protons}
\label{app:spec}
We describe the escaping proton population by the differential injection rate $Q'_p(E'_p,\Omega')$ in the jet comoving frame. The distribution is assumed to be isotropic, with its normalization determined by $L'_p/2=\int dE'_p\,d\Omega'\,E'_p Q'_p(E'_p,\Omega')$.
Escape occurs through a surface fixed in the observer frame. For two escape events at the same orbital height, separated by $dt$ in the observer frame, the corresponding time interval in the comoving frame is $dt'=\Gamma_{\rm j}dt$. The conservation of particles requires that $Q_{p,\xi}(E_p,\Omega)\,dE_p\,d\Omega\,dt = Q'_p(E'_p,\Omega')\,dE'_p\,d\Omega'\,dt'$, thus we have:
\begin{equation}
Q_{p,\xi}(E_p,\Omega)
=
\Gamma_{\rm j}\delta_\xi
Q'_p(\frac{E_p}{\delta_\xi},\Omega').
\end{equation}
The power for the escaping protons in the observer frame is then obtained by
\begin{equation}
L_p=\sum_{\xi=\pm1}\int dE_p\,d\Omega\,E_p Q_{p,\xi}(E_p,\Omega)\,,
\end{equation}
with equal intrinsic proton powers in the two branches. This gives $L_p=\Gamma_{\rm j}^2L'_p$.

\section{Calculation details and model checks}
\label{app:calc}

Table~\ref{tab:fit} lists the priors and marginalized constraints used in the joint fit.

\begin{table*}[t]
\caption{Priors and posterior summaries for the joint fit}
\label{tab:fit}
\begin{ruledtabular}
\begin{tabular}{lcccccc}
Parameter &
$H_{\rm GeV}/10^{11}{\rm cm}$ &
$\beta_{\rm j}$ &
$\theta_{\rm j}$ [deg] &
$\varphi_{\rm j}$ [deg] &
$p_e$ &
$K_{\rm inj}/10^{48}{\rm s^{-1}}$ \\

Prior &
$[1,30]$ &
$[0.01,0.99]$ &
$[0,90]$ &
$[0,360]$ &
$[3,5]$ &
$[10^{-3},10^2]$ \\

Best fit ($1\sigma$) &
$2.751^{+0.327}_{-0.600}$ &
$0.603^{+0.076}_{-0.071}$ &
$44.97^{+10.71}_{-3.50}$ &
$196.32^{+6.08}_{-2.25}$ &
$3.639^{+0.135}_{-0.061}$ &
$0.995^{+1.390}_{-0.258}$ \\

\hline

Parameter &
$H_{\rm esc}/10^{10}{\rm cm}$ &
$\beta_{\rm DW}$ &
$L'_p(>1\,{\rm TeV})/10^{38}{\rm erg\,s^{-1}}$ &
$s_p$ &
$E'_{p,\rm cut}$ [PeV] &
\\

Prior &
$[0.1,30]$ &
$[0.03,0.40]$ &
$[0.1,10]$ &
$[2,3]$ &
$[10,10^3]$ &
\\

Best fit ($1\sigma$) &
$2.971^{+3.026}_{-2.699}$ &
$0.142^{+0.187}_{-0.015}$ &
$0.583^{+2.574}_{-0.355}$ &
$2.461^{+0.213}_{-0.165}$ &
$107.78^{+439.66}_{-80.48}$ &
\\
\end{tabular}
\end{ruledtabular}

\vspace{0.5ex}
\noindent
\end{table*}

\textit{Radiation field, beaming, and attenuation.--}
We approximate the WR star as a blackbody with radius $R_\star=10^{11}\,$cm and temperature $T_\star=1.0\times10^5\,$K \cite{2007ARA&A..45..177C,Koljonen2017,Vilhu2021}. Its differential photon density at the stellar surface is
\begin{equation}
n_\star(\epsilon,\Omega_\star)=\frac{2}{(hc)^3}\frac{\epsilon^2}{\exp(\epsilon/k_BT_\star)-1}.
\label{eq:end_bb}
\end{equation}
The finite angular size of the star is retained in Eq.~(\ref{eq:pgamma}). At proton position $\boldsymbol r_{p,\xi}$, define $R=|\boldsymbol r_{p,\xi}|$, $\mu_{0,\xi}=\boldsymbol e_{\rm obs}\!\cdot\!\boldsymbol r_{p,\xi}/R$, and angular coordinates $(\omega,\varphi)$ on the apparent stellar disk. The collision angle is
\begin{equation}
\begin{split}
\mu_{p,\xi}=\mu_{0,\xi}\cos\omega+\sqrt{1-\mu_{0,\xi}^2}\sin\omega\cos\varphi,\,
\\ \omega \in \left[0, \arcsin(R_\star/R)\right].
\label{eq:end_mu}
\end{split}
\end{equation}
The $p\gamma$ photon yield in Eq.~(\ref{eq:pgamma}) is calculated with \texttt{SOPHIA++} \cite{Mucke2000,ABR2022}, using spectra averaged over $2\times10^4$ simulated interactions for each interaction energy. The observed flux at $d=9\,$kpc is then written as
\begin{equation}
\frac{dF_\gamma}{dE_\gamma}=\frac{1}{d^2}\sum_{\xi=\pm1}
\left(\frac{dN_{\gamma,\xi}^{p\gamma}}{dE_\gamma dt}+\frac{dN_{\gamma,\xi}^{pp}}{dE_\gamma dt}\right).
\label{eq:end_flux}
\end{equation}
The emitted gamma rays are attenuated by pair production in the finite stellar radiation field, for which we integrate
\begin{equation}
\frac{d\tau_{\gamma\gamma,\xi}}{dl}=\int d\epsilon\,d\Omega_\star\,n_\star(\epsilon,\Omega_\star,\boldsymbol r_{\gamma,\xi})(1-\mu_{\gamma,\xi})\sigma_{\gamma\gamma}(E_\gamma,\epsilon,\mu_{\gamma,\xi}),
\end{equation}
and by the CMB and Galactic interstellar radiation field during propagation \cite{1967PhRv..155.1404G,1967PhRv..155.1408G,2009ApJ...707..916F,2017MNRAS.470.2539P}.

\textit{Circum-binary wind environment.--}
Cygnus~X-3 undergoes pronounced X-ray/radio state changes, and its large X-ray polarization supports an optically thick outflow associated with supercritical accretion \cite{Koljonen2017,Vilhu2021,Veledina2024}. In the fiducial calculation both winds are taken to be freely expanding, with effective nucleon density
\begin{equation}
n_i(r_i)=\frac{\dot M_i}{4\pi r_i^2v_i m_p},\qquad i={\rm WR,DW}.
\label{eq:end_wind_density}
\end{equation}
For the WR wind we use $\dot M_{\rm WR}=5\times10^{-6}M_\odot\,{\rm yr^{-1}}$ and $v_{\rm WR}=1000\,{\rm km\,s^{-1}}$, giving $n_{\rm WR}\simeq1.8\times10^{12}(r/a)^{-2}\,{\rm cm^{-3}}$. The disk-wind mass-loss rate is not directly measured. Taking $L_X\sim10^{39}\,{\rm erg\,s^{-1}}$ and $L_X=\eta_{\rm acc}\dot M_{\rm in}c^2$ gives $\dot M_{\rm in}\simeq1.8\times10^{-7}(\eta_{\rm acc}/0.1)^{-1}M_\odot\,{\rm yr^{-1}}$. In the supercritical picture, most of the excess supply inside the spherization radius is expelled, so that
\begin{equation}
\dot M_w\simeq\dot M_{\rm supply}-\dot M_{\rm in}\sim(\dot m_{\rm acc}-1)\dot M_{\rm in}.
\end{equation}
For $\dot m_{\rm acc}\sim10$--$100$, this gives a wind loss of order $10^{-6}$--$10^{-5}M_\odot\,{\rm yr^{-1}}$ \cite{King2016}; we therefore adopt $\dot M_{\rm DW}=10^{-6}M_\odot\,{\rm yr^{-1}}$. The same model gives
\begin{equation}
v_w\simeq\left(\frac{2l'\eta_{\rm acc}}{\dot m_w}\right)^{1/2}c,
\end{equation}
where $\dot m_w\simeq\dot m_{\rm acc}-1$
, and $l'$ is the effective wind-driving luminosity in Eddington units ($l'\lesssim 1+\ln\dot m_{\rm acc}$). For supercritical accretion, mildly relativistic winds are expected. Our fit gives $v_{\rm DW}\simeq0.14c$ and $n_{\rm DW}\simeq 8.4 \times10^9(r/a)^{-2}\,{\rm cm^{-3}}$.

The interaction of the two winds is specified by $\eta_w=\dot M_{\rm DW}v_{\rm DW}/(\dot M_{\rm WR}v_{\rm WR})$. In the isotropic fiducial model the contact discontinuity (CD) follows the thin-shell solution \cite{Canto1996},
\begin{equation}
R_{\rm CD}=\frac{a\sin\theta_{\rm DW}}{\sin(\theta_{\rm WR}+\theta_{\rm DW})}.
\label{eq:end_cd}
\end{equation}
and
\begin{equation}
    \theta_{\rm DW}\cot\theta_{\rm DW}=1+\eta_w^{-1}(\theta_{\rm WR}\cot\theta_{\rm WR}-1),
\end{equation}
where $\theta_{\rm WR}$ and $\theta_{\rm DW}$ are the polar angles of a contact-surface point viewed from the WR star and the compact object, respectively. At every point along the proton trajectory, we determine which side of the CD the proton lies on by comparing $|r_p|$ and $R_{\rm CD}$. Shocked-layer structure, mixing, and clumping are neglected. We also neglect the target-wind angular correction in $pp$ collisions. Even for $\beta_{\rm DW}\simeq0.2$, the relative-flux factor $1-\beta_{\rm DW}\mu$ spans only $\simeq0.8$--$1.2$, and the collisions are far above pion-production threshold; this contrasts with $p\gamma$, for which the angular factor determines the stellar photons above photopion threshold.

\textit{Finite-opening-angle disk wind.--}
The fiducial isotropic disk wind is a computational simplification. We also consider a bipolar top-hat outflow aligned with the jet axis,
\begin{equation}
n_{\rm DW}(r,\hat{\boldsymbol r})=\frac{\dot M_{\rm DW}}{4\pi r^2v_{\rm DW}m_p}f_{\rm DW}(\hat{\boldsymbol r}),
\label{eq:end_aniso}
\end{equation}
where
\begin{equation}    
f_{\rm DW}=\begin{cases}
(1-\cos\theta_w)^{-1},&|\hat{\boldsymbol r}\!\cdot\!\boldsymbol e_j|\geq\cos\theta_w,\\
0,&|\hat{\boldsymbol r}\!\cdot\!\boldsymbol e_j|<\cos\theta_w,
\end{cases}
\end{equation}
which conserves the total mass-loss rate over the two cones. The angle-dependent momentum flux is included by integrating its transverse and axial components,
\begin{align}
\dot P_{{\rm DW},\perp}&=\frac{\dot M_{\rm DW}v_{\rm DW}}{4\pi}\int_0^{\theta_{\rm DW}}f_{\rm DW}\sin^2\theta\,d\theta, \\
\dot P_{{\rm DW},\parallel}&=\frac{\dot M_{\rm DW}v_{\rm DW}}{4\pi}\int_0^{\theta_{\rm DW}}f_{\rm DW}\sin\theta\cos\theta\,d\theta,\\
\dot P_{{\rm WR},\perp}&=\frac{\dot M_{\rm WR}v_{\rm WR}}{4\pi}\left(\frac{\theta_{\rm WR}}{2}-\frac{\sin2\theta_{\rm WR}}{4}\right), \\
\dot P_{{\rm WR},\parallel}&=\frac{\dot M_{\rm WR}v_{\rm WR}}{4\pi}\frac{\sin^2\theta_{\rm WR}}{2},
\end{align}
and the contact surface is obtained from
\begin{equation}
\frac{r_\parallel}{a}\dot P_\perp-\frac{r_\perp}{a}\dot P_\parallel-\dot P_{{\rm WR},\perp}=0.
\end{equation}
Here $r_\parallel=r\cos\theta_{\rm DW}$ and $r_\perp=r\sin\theta_{\rm DW}$, with $\dot P_\perp=\dot P_{{\rm DW},\perp}+\dot P_{{\rm WR},\perp}$ and $\dot P_\parallel=\dot P_{{\rm DW},\parallel}-\dot P_{{\rm WR},\parallel}$. An illustrative configuration with $\theta_w\simeq32^\circ$, $\dot M_{\rm DW}=4\times10^{-7}M_\odot\,{\rm yr^{-1}}$, and $v_{\rm DW}\simeq0.25c$ reproduces the observed UHE spectrum and orbital modulation with the other fit parameters nearly unchanged. Thus the low target density required along the present observer-directed proton trajectories does not rely on a globally isotropic wind. If the wind axis follows the inner flow or jet, a change of orientation can move those trajectories out of the disk-wind cone and into the denser WR wind without a comparable reduction in accretion or proton power.

\textit{Pure WR-wind limit and comparison with Ref.~\cite{Zdziarski2026_PeV}.--}
To isolate the environmental effect, we also remove the disk wind and set $n_{\rm gas}=n_{\rm WR}$. The $p\gamma$ efficiency is unchanged, whereas the $pp$ efficiency rises by roughly two orders of magnitude. To reproduce the present LHAASO flux in this limit we use, as in Ref.~\cite{Zdziarski2026_PeV}, $s_p=2.2$ and $L_p(>1\,{\rm TeV})\simeq2\times10^{36}\,{\rm erg\,s^{-1}}$ without a high-energy exponential cutoff. The employed $s_p$ is harder than the best-fit value in our fiducial model and may reflect a change in the accelerated proton spectrum between different accretion/jet states.
We can reproduce the spectrum below $1\,$PeV, while the multi-PeV $p\gamma$ bump becomes negligible; the $pp$ component is phase dependent because the WR-wind column varies with orbital phase. Conversely, adopting a proton power comparable to the fiducial joint fit raises the $pp$ flux by hundreds relative to the current LHAASO level, approaching the historical PeV range shown in Fig.~\ref{fig:historical}.

For comparison, Fig.~\ref{fig:historical} also shows the geometry adopted by Ref.~\cite{Zdziarski2026_PeV}, $\theta_j=25^\circ$, $\varphi_j=120^\circ$, and $H_{\rm esc}=0.7a$, with an isotropic distribution of escaping protons in the observer frame. We successfully reproduce their UHE result. In this geometry, the counterjet contribution dominates. However, the same geometry does not reproduce the contemporaneous GeV orbital modulation, illustrating the additional constraint supplied by the joint GeV--PeV modeling.

\end{document}